\documentclass[letterpaper, 10 pt, conference]{ieeeconf}  % Comment this line out if you need a4paper

\IEEEoverridecommandlockouts                              % This command is only needed if 
\usepackage{graphics} % for pdf, bitmapped graphics files
\usepackage{epsfig} % for postscript graphics files

\title{\LARGE \bf
Android dialogue system for customer service using prompt-based topic control and compliments generation
}

\author{Tamotsu Miyama$^{1}$ and Shogo Okada$^{1}$% <-this % stops a space
\thanks{$^{1}$1
Japan Advanced Institute of Science and Technology,
{\tt\small tamotsu.miyama@jaist.ac.jp},{\tt\small okada-s@jaist.ac.jp}}%
}

\begin{document}

\maketitle
\thispagestyle{empty}
\pagestyle{empty}

%%%%%%%%%%%%%%%%%%%%%%%%%%%%%%%%%%%%%%%%%%%%%%%%%%%%%%%%%%%%%%%%%%%%%%%%%%%%%%%%
\begin{abstract}

%This paper describes a dialogue system developed for the Dialogue Robot Competition 2023, which uses the ChatGPT-API to achieve topic control by inserting text into prompts. 
This paper describes a dialogue system developed for the Dialogue Robot Competition 2023 that achieves topic control for trip planning by inserting text into prompts using the ChatGPT-API.
We built a system that is capable of generating compliments for the user based on recognition of the user's appearance and creating travel plans by extracting the knowledge about the user's preference from the history of the user's utterances.
Complements and planning based on preference are the elements required to maintain the quality of customer service. 
A preliminary round was held at a travel agency's actual store, where real customers experienced and evaluated the system. This system was evaluated first in the preliminary round and participated in the final round.  The results of the preliminary round showed the effectiveness of the proposed system.
\end{abstract}

%%%%%%%%%%%%%%%%%%%%%%%%%%%%%%%%%%%%%%%%%%%%%%%%%%%%%%%%%%%%%%%%%%%%%%%%%%%%%%%%
\section{INTRODUCTION}
% (私の修正文)
This paper describes the dialogue system we developed for the Dialogue Robot Competition 2023 \cite{c1}\cite{c2}: the advent of generative AI such as ChatGPT has dramatically improved the performance of response generation  of dialogue systems \cite{c3}. In addition of appropriate dialogue ability,  hospitality is required in customer service such as travel agencies, and dialogue systems for customer service are expected to response with hospitality \cite{c4}. Furthermore, from the aspect of social implementation of robotic dialogue systems, it is important to conduct demonstration tests in actual fields and extract issues. In this context, it is necessary to construct a dialogue system with various elements of hospitality service, and to evaluate users.
%This paper describes the dialogue system we developed for the Dialogue Robot Competition 2023 \cite{c1}\cite{c2}: the advent of generative AI such as ChatGPT has dramatically improved the accuracy of dialogue systems \cite{c3}. In addition, hospitality is required in customer service such as travel agencies, and dialogue systems with this spirit are needed \cite{c4}. Furthermore, from the aspect of social implementation of robotic dialogue systems, it is important to conduct demonstration tests in actual fields and extract issues. In this context, it is necessary to construct a dialogue system with various elements of hospitality service and to evaluate users.

There are three key points in the development of the proposed dialogue system. (1) Topic control for trip planning was achieved by using ChatGPT for dialog control and inserting text in the prompts. 
%(2) Customer service requires various elements such as language, facial expressions, listening, and attitude, and we were able to understand the user's external characteristics and issue compliments \cite{c4}. 
(2) Customer service requires various elements such as language, facial expressions, listening, and attitude \cite{c4}, as well as the use of compliments to establish a personal relationship with the customer \cite{c5}. Therefore, this system was designed to identify the user's external characteristics and issue compliments.
%Since the (3) target task was to create a travel plan to visit two tourist attractions in Kyoto, 
(3) In order to propose travel plans that match the user's preferences, we implemented a function to create a travel plan using the user's past utterances. 
In the preliminary round, experiments and evaluations were conducted with customers who actually visited the travel agency's stores. We placed first out of 19 teams entered (12 teams participated) and were selected to participate in the final round.

\section{Proposed System}

\begin{figure}[hbtp]
 \centering
 \includegraphics[keepaspectratio, scale=0.32]
      {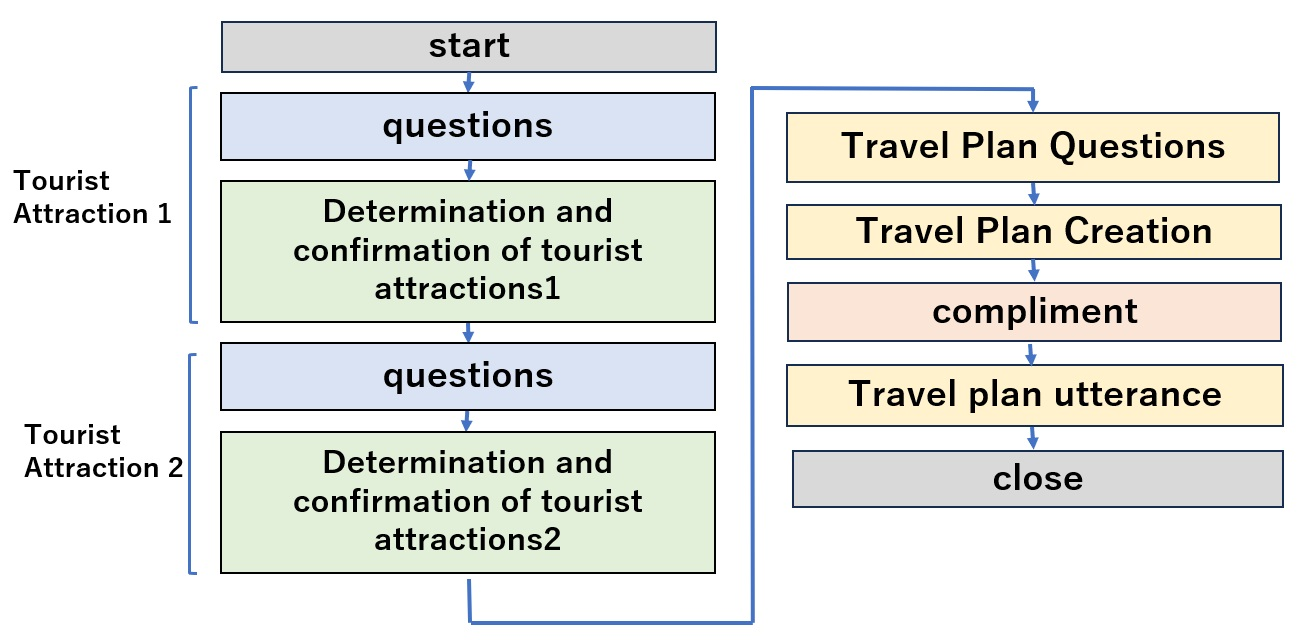}
 \caption{dialogue flow}
 \label{dialogue_flow}
\end{figure}

\subsection{Controlling topics with ChatGPT prompts}

%The generative AIs used were GPT-3.5-trubo and GPT4,
The generative AIs used were GPT-3.5-trubo and GPT4.
%which kept a dialogue history in the prompts and controlled the dialogue and topics by inserting text into the prompts. 
Since the goal task is to create a travel plan, insert fixed text in the prompt to direct the topic toward that task.
%The three basic roles given to the prompt are "system," "assistant," and "user. The "system" controls the dialogue by giving instructions regarding the next comment to be uttered. The "assistant" controlled the next topic of conversation by inserting question text in the middle of the prompt.

\subsection{Dialogue Flow}

The dialogue flow is shown in Figure \ref{dialogue_flow}. Basically, the customer's requests are elicited by asking questions, and then tourist destinations are determined one by one. The customer's requirements for the travel plan are then confirmed, and a travel plan that meets the customer's requirements is then discussed.

The system asked the customer questions about their experience visiting Kyoto, enjoyment of the trip, and desired activities, and the system referred the answers to the ChatGPT to determine Sightseeing Spot 1. The system asked the customer questions about what they wanted to eat, their preferred areas, hobbies, etc., and the answers were queried on ChatGPT to determine Sightseeing Spot 2. In determining the sightseeing spots, the existence of data was checked by referring to the Sightseeing Spot API. In addition, the summary description of the sightseeing spot being uttered was created by having ChatGPT summarize the introduction in the sightseeing spot API. In the process of creating a travel plan, we asked the customer about the place of departure and means of transportation and used the answers to create a travel plan to visit the two sightseeing spots via ChatGPT.

\subsection{Function to complement a user's physical appearance}

\begin{figure}[hbtp]
 \centering
 \includegraphics[keepaspectratio, scale=0.23]
      {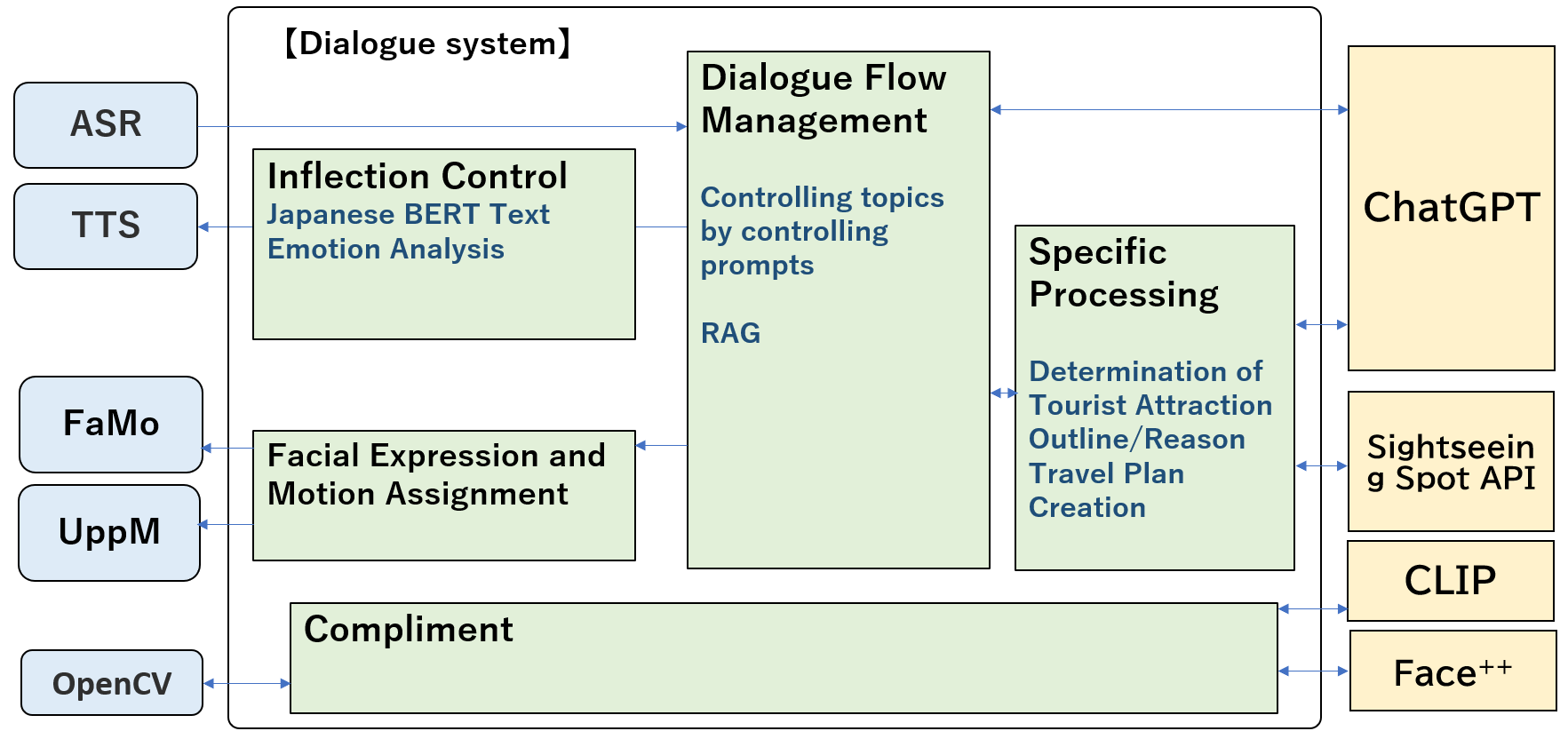}
 \caption{Overall Configuration.
 FaMo and UppM denote facial expression generation and mouth shape generation and upper body position generation}
 \label{overall_configuration}
\end{figure}

A dialogue system designed for customer service requires various elements such as language, facial expressions, gestures, and listening, among which this system focuses on words of praise.

The specific words of praise for each user are automatically determined based on the user's appearance recognition as follows. 
During the dialogue, images of the user's upper body was captured with a camera, and the user's appearance characteristics were extracted from this image. 

The color of clothing, the shade of clothing are recognized from the image capturing user's appearance,
by using CLIP model, which is a pre-trained image classification model developed by OpenAI.
In addition to the appearance information,
Gender type, age, glasses, and beauty quotient are estimated using Face++.

%CLIP is a pre-trained image classification model developed by OpenAI, which can classify images for a wide range of tasks without fine-tuning, and
%Face++ is an image recognition platform provided by Megvii, Inc. CLIP[N] recognizes various features of facial images, such as the color of clothing, the shade of clothing, and glasses,
%Face++ obtained gender, age, glasses, and beauty quotient. 

A maximum of two compliments were generated based on the estimated five characteristics: clothing color, clothing shade, eyeglasses, beauty, and personality. Specifically, "Your chic clothes look great on you!" You are such a sweet person!" and other comments.

\subsection{Control using user's past speech}

ChatGPT was also used to determine sightseeing spots and to create travel plans. In addition, ChatGPT was queried using the user's past speech information for these decisions and creation. For example, in the case of creating a travel plan, the user was asked about companions, interests, and activities in the first half of the dialogue flow, and the user's responses were included in the prompts and queried to ChatGPT to create a travel plan.

\subsection{Overall Configuration}

The overall system configuration is shown in Figure \ref{overall_configuration}. 
%The main process is to manage topics and the entire dialogue flow by controlling ChatGPT prompts in the dialogue flow management module. We also built modules for inflection, facial expressions and actions, control related to praise, and specific processing such as travel plan creation.

\section{User Evaluation and Preliminary Results}

\begin{figure}[hbtp]
 \centering
 \includegraphics[keepaspectratio, scale=0.35]
      {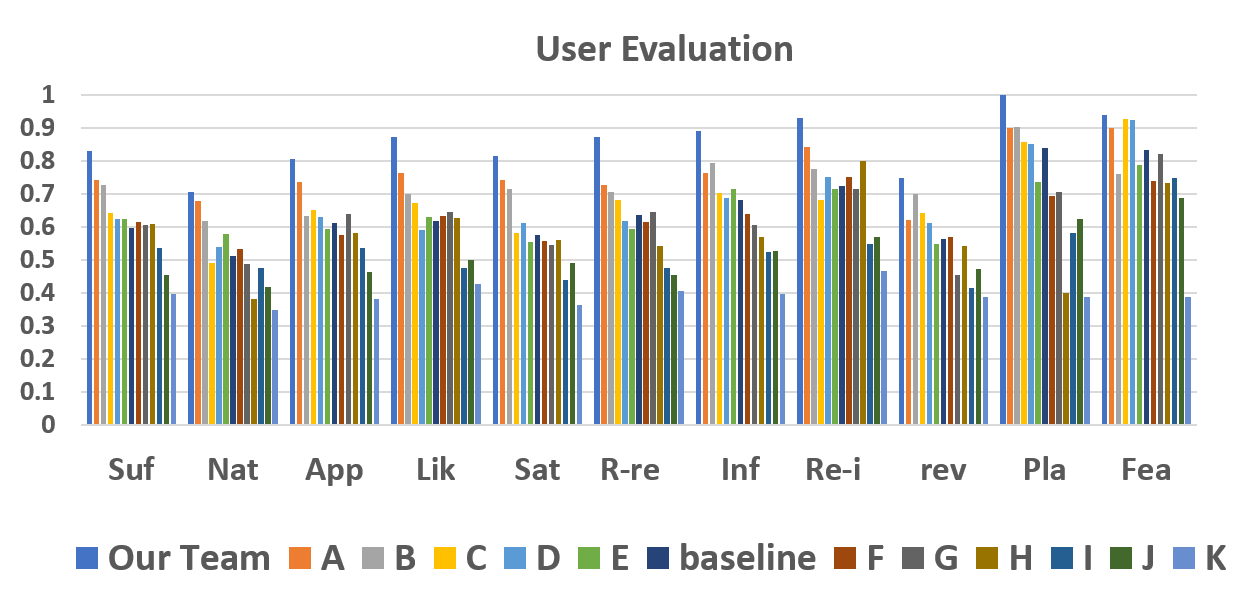}
 \caption{User Evaluation and Preliminary Results. 
  Suf, Nat, App, Lik, Sat, R-re, Inf, Re-i, Rev, Pla, and Fea denote Sufficiency of information, Naturalness, Appropriateness, Likability, Satisfaction, Robot reliability, Information reference, Reliability of information, Degree of revisit, Plan creation, and Feasibility of plan.}
 \label{User_Eva}
\end{figure}

The evaluation items are largely divided into two categories: satisfaction evaluation of the experiencer and evaluation of the plan. The satisfaction evaluation consists of 9 items (7 on each recoded scale): sufficiency of information, naturalness of dialogue, appropriateness of dialogue, friendliness of response, satisfaction with dialogue, trustworthiness of the robot, informational reference, reliability of information, and degree of desire to return, all of which were rated No. 1. The plan evaluation item, which asks whether a plan to visit two sightseeing spots can be created (Yes/No) and whether it is feasible (Yes/No), was also rated No. 1. Figure \ref{User_Eva} shows the satisfaction rating and the plan evaluation on a plan diagram. Therefore, of course, the overall evaluation was also the first place in the preliminary round. This means that our dialogue system was the highest rated in the evaluation by real customers in the actual shops. In addition, the high evaluation of the reliability of information in the satisfaction ratings suggests that control using the user's past speech contributed significantly to improving the user's reliability.

\section{CONCLUSION}

I described our dialogue robot developed for the Dialogue Robot Competition 2023. Our dialogue system used ChatGPT to control the dialogue and topics by embedding text in the prompts as appropriate. We also implemented a function to praise the user's physical appearance and a function to create a travel plan using the user's previous utterances. As a result of the preliminary round evaluation, the system was ranked first in both satisfaction rating and plan rating.

\section*{ACKNOWLEDGMENT}

%We would like to thank the Organizers of the Dialogue Robot Competition 2023 for providing the necessary software, androids, experimental environment, and various opportunities for academic research to build this dialogue system.
We deeply appreciate the organizers of the Dialogue Robot Competition 2023.

%%%%%%%%%%%%%%%%%%%%%%%%%%%%%%%%%%%%%%%%%%%%%%%%%%%%%%%%%%%%%%%%%%%%%%%%%%%%%%%%

\end{document}